\documentclass[nofootinbib,aps,10pt,twocolumn]{revtex4-1}

\usepackage{array,amsmath,amsthm}
\usepackage{mathtools}
\usepackage{graphicx}
\usepackage{color}
\usepackage{amssymb}

\begin{document}

\hfill
\begin{minipage}{20ex}\small
ACFI-T17-05\\
ZTF-EP-17-04
\end{minipage}

\title{On the Decoupling Theorem for Vacuum Metastability}
\author{Hiren H. Patel}
\email{hhpatel@umass.edu}
\affiliation{Amherst Center for Fundamental Interactions, Department of Physics,\\ University of Massachusetts, Amherst, MA 01003}
\author{Branimir Radov\v{c}i\'{c}}
\email{radovcic@mpi-hd.mpg.de}
\affiliation{Max-Planck-Institut f\"{u}r Kernphysik\\
Saupfercheckweg 1, 69117 Heidelberg, Germany}
\affiliation{Department of Physics, Faculty of Science, University of Zagreb \\
Bijeni\v{c}ka cesta 32, 10000 Zagreb, Croatia}


\begin{abstract}
In this paper, we numerically study the impact heavy field degrees of freedom have on vacuum metastability in a toy model, with the aim of better understanding how the decoupling theorem extends to semiclassical processes.  We observe that decoupling applies to partial amplitudes associated with fixed final state field configurations emerging from the tunneling processes, characterized by a scale such as the inverse radius of a spherically symmetric bubble, and not directly on the total lifetime (as determined by the ``bounce'').  More specifically, tunneling amplitudes for bubbles with inverse radii smaller than the scale of the heavier fields are largely insensitive to their presence, while those for bubbles with inverse radii larger than that scale may be significantly modified. 
\end{abstract}

\maketitle

\section{Introduction}

Recently, Branchina \cite{Branchina:2013jra, Branchina:2014usa, Branchina:2014rva} has observed that in contrast to the perturbative contributions to a physical process derived from an effective theory conforming to the decoupling theorem \cite{Appelquist:1974tg}, non-perturbative tunneling contributions may exhibit much greater sensitivity to the scale of new physics than intuition would suggest.

The observation was made in the context of electroweak vacuum metastability, where additional higher-dimensional operators added to the Coleman-Weinberg effective potential parameterizing new physics near the Planck scale appeared to increase the zero-temperature tunneling rate by over 700 orders of magnitude relative to the Standard Model rate \cite{Isidori:2001bm,EliasMiro:2011aa,Degrassi:2012ry}.  This effect was subsequently confirmed in \cite{DiLuzio:2015iua,Andreassen:2016cvx}, and can be traced down to the modification of the bounce solution that is much smaller with field values reaching the Planck scale.  Although we do not dispute the effect, since the field value in the center of the bounce solution reaches values on the order of $\Lambda_\text{UV}$, we believe that the analysis is somewhat inconsistent from the effective field theory (EFT) point of view.  More specifically, the finite set of operators alone no longer appropriately parametrize new physics appearing at that scale.  Additionally, we find it concerning from the standpoint of the decoupling theorem, where intuition suggests that the addition of new physics should not significantly affect rates at the low scale.

We revisit this phenomenon to improve our conceptual understanding of how decoupling operates for semiclassical processes in a toy $\phi^4$ theory.  To avoid the inconsistency associated with a naive EFT parametrization, we couple the theory to a heavy scalar or to a heavy fermion as representative models of short distance physics.  To consistently capture the effect of heavy physics on tunneling, we work with a more complete functional form of the action approximated by an expansion in the coupling constant.  We find that the addition of a fermion with a sufficiently large coupling constant can significantly modify tunneling rates, as originally observed by Branchina.

Below, we argue that vacuum tunneling is not necessarily a low energy process, and therefore decoupling does not necessarily apply to the total rate.  Rather, it applies to partial amplitudes associated with fixed final state field configurations emerging from the tunneling processes.  As a result, one should not expect the total tunneling rate of the metastable vacuum to be insensitive to new physics.

\section{Particle decay}
To motivate this discussion, we illustrate in a hypothetical scenario how new physics could have a drastic effect in the more familiar process of neutron decay:
\begin{gather*}
\begin{aligned}
\Gamma_1: & \qquad n \rightarrow p e^- \bar\nu\\
\Gamma_2: & \qquad n \rightarrow \pi^0 \bar\nu
\end{aligned}
\end{gather*}
The neutron initial state represents the unstable electroweak vacuum in our analogy, while the individual modes of decay correspond to two possible emerging field configurations in a tunneling event.  The first listed channel is the familiar neutron beta decay with a $Q$-value of 0.782 MeV.  The second channel is the forbidden $B$-violating process with a much larger $Q$-value of 805 MeV.  The total width of the neutron is given by the sum of partial widths $\Gamma = \Gamma_1 + \Gamma_2$.  In the low energy theory, the total width is dominated by $\Gamma_1$.

Consider the addition of a heavy particle, representing new physics, of mass $M = 500\text{ MeV}$ that couples to both channels.
How would the total lifetime of the neutron be altered by this new degree of freedom?  Although the contribution to $\Gamma_1$ would be suppressed by $Q^2/M^2 \sim 10^{-6}$, it would be incorrect to conclude that the total width $\Gamma$ would be insensitive to new physics.  The second channel has a much larger $Q$-value, and the contributing virtualities would sample the presence of new physics, effectively generating a very large $\Gamma_2$.  As a result, the lifetime of the neutron in our hypothetical example would be significantly shortened by the presence of new physics.

Applying this to the problem of vacuum metastability, the presence of new physics at a scale $\Lambda$ may magnify the ``high energy'' partial width tunneling amplitudes which were small before the addition of new physics.  Below, we show how this happens in a toy $\phi^4$ theory.

\section{Low energy theory}
Throughout this study, our low energy theory will be the real scalar $\phi^4$ theory, with the potential
\begin{equation}\label{eq:lowEnergyPot}
V(\phi) = \frac{1}{2} m_\phi^2 \phi^2 - \eta_\phi \phi^3 + \frac{1}{8} \lambda_\phi \phi^4
\end{equation}
suitably modified to exhibit an instability.  We will consider this theory in two separate cases.  In the first case, which we call the ``asymmteric double well potential" the low energy constants $m_\phi^2$, $\eta_\phi$ and $\lambda_\phi$ are all taken positive.  In the second case, which we call the ``unbounded potential", we take $\lambda_\phi$ negative and, for simplicity, $\eta_\phi=0$.  In both cases, the low energy metastable phase is at $\phi=0$.

\section{Evaluation of Partial Tunneling Amplitudes}\label{sec:PartialAmplitude}
Instead of the full width as calculated semiclassically by methods developed originally by Coleman and Callan \cite{Coleman:1977py,Callan:1977pt}, we are interested in the effect of new physics on amplitudes for exclusive final states, corresponding to specific profiles of the field that emerges upon tunneling.  This is because we would like to study the sensitivity of new physics on these amplitudes separately.  Technically, we expect that upon a consistent evaluation of such amplitudes, the sum over the corresponding partial widths should yield the full width that matches the method of Coleman and Callan.  However we are not aware of a method in the literature to compute these amplitudes\footnote{A preliminary formalism has been outlined in \cite{Dine:2012tj} in the context of studying the effect of Lorentz transformation of tunneling rates.}.  Fortunately, we will not need the full machinery for the careful evaluation of partial widths.  Instead we will be content to investigate just the representative contributions to an exclusive amplitude, which we summarize here.

We are interested in calculating the amplitude for the system to make a transition from the false vacuum $\phi_\text{FV} = 0$ at time $t_i\rightarrow-\infty$ to a specified final state $\phi_\text{f}$ at time $t_\text{f}$.  The Feynman path integral representation of this amplitude is
\begin{gather}
\langle \phi_\text{f}(\textbf{x})\vert e^{-i H (t_\text{f}-t_\text{i})} \vert \phi_\text{FV}(\textbf{x}) \rangle\ = \int_{\phi_\text{FV}}^{\phi_\text{f}} \mathcal{D} \phi \, e^{i S[\phi(x)]}\,,\\
S[\phi(t,\textbf{x})] = \int_{t_\text{i}}^{t_\text{f}} dt \int d^3 \textbf{x} \, \mathcal{L}[\phi(t,\textbf{x})]\,,\\
\label{eq:lagrangian}\mathcal{L}[\phi(t,\textbf{x})] = \frac{1}{2} \Big(\frac{d \phi}{d t}\Big)^2 - \frac{1}{2} (\nabla \phi)^2 - V(\phi)\,.
\end{gather}
A proper evaluation of this amplitude in the stationary phase approximation would require one to solve a partial differential equation with insufficient symmetry to reduce it to an ordinary differential equation.  
To make analytic progress we shall compute a representative contribution to this amplitude by transforming the field theoretic problem to a one dimensional quantum mechanical problem\footnote{For a similar idea used to analyze electroweak sphaleron transitions, \cite{Tye:2015tva}.} by restricting the integral to a single family of paths parametrized by one dynamical coordinate $f(t)$.  This is arranged by fixing the spacial field profile up to one free dynamical coordinate $f(t)$ at each point in time.

In what follows, we will use the family of Gaussian bubbles
\begin{equation}\label{eq:gaussianfamily}
\phi_\text{G}(t,\textbf{x})=f(t) e^{- r^2/R^2}\,,
\end{equation}
dependent upon the dynamical coordinate $f(t)$, and a scale parameter $R$ which will be related to the specific final state for the tunneling process.  The dynamical coordinate satisfies $f(t\rightarrow-\infty) = 0$ corresponding to the false vacuum as the initial state $\phi = \phi_\text{FV} \equiv 0$, and $f(t_\text{f}) = f_\text{f}$ corresponding to the emerging bubble as the final state
\begin{equation}\label{eq:finalbubble}
\phi_\text{f} = f_\text{f}e^{-r^2/R^2}.
\end{equation}
The scale parameter $R$ and the field value at the center of the final state bubble $f_\text{f}$  are connected by energy conservation
\begin{equation}
E[\phi_\text{f}(\textbf{x})] = \int d^3 \textbf{x} [\frac{1}{2} (\nabla \phi_\text{f}(\textbf{x}))^2 + V(\phi_\text{f}(\textbf{x}))] = 0 \,,
\end{equation}
and ultimately fixes $f_\text{f} \sim R^{-1}$.  Although the precise form we take for the family of field configurations is not crucial to our analysis, we emphasize that the parameter $R^{-1}$ which sets the scale of the final state bubble is like the $Q$-value of the particle decay analogy of the previous section.  That is, we will find that tunneling processes for large $R$ is like the particle decay process with small $Q$-value and is insensitive to new physics, while those that tunnel to small $R$ are like particle decay processes with large $Q$-value making them more sensitive to new physics.  This is not surprising since Fourier modes of the field profile are peaked at $f_\text{f}/R$.

From the field theory Lagrangian in (\ref{eq:lagrangian}), we obtain the reduced Lagrangian for the dynamical variable $f$
\begin{multline}
L_\text{R}[f(t)] = \frac{1}{2} \Big(\frac{d f}{d t}\Big)^2 \frac{\pi^{3/2} R^3}{2 \sqrt{2}} -  \frac{1}{2} f^2 \frac{3 \pi^{3/2} R}{2 \sqrt{2}} \\
- \int d^3 \textbf{x} \, V(f e^{- r^2/R^2})\,.
\end{multline}
We achieve canonical normalization for the kinetic term by making the change of variables $t = \frac{\pi^{3/2} R^3}{2 \sqrt{2}} t_\text{R}$ yielding
\begin{equation}
L_\text{R}[f(t_\text{R})] = \frac{1}{2} \Big(\frac{d f}{d t_\text{R}}\Big)^2 -  U(f)\,,
\end{equation}
where the reduced potential for $f$ is
\begin{equation}\label{eq:ReducedPotential}
U(f)= \frac{3 \pi^3 R^4}{16} f^2 + \frac{\pi^{3/2} R^3}{2 \sqrt{2}} \int d^3 \textbf{x} \, V(f e^{- r^2/R^2}) \,.
\end{equation}
Using this action, we can compute the tunneling amplitude in the WKB approximation,
\begin{equation}
A_{f_\text{f}} \sim e^{- \int_{f_\text{i}}^{f_\text{f}} \sqrt{2 U(f)} df} \,,
\end{equation}
subject to
\begin{equation}
U(f_\text{i})=U(f_\text{f})=0\,.
\end{equation}
In this picture, the reduced potential $U$ can be understood as the one the system has to effectively tunnel through to emerge as the profile given in (\ref{eq:finalbubble}), and therefore depends on $R$.

\section{Introducing New physics}
We would like to avoid characterizing the effect of new physics by a limited set of high dimensional operators for consistency reasons explained in the introduction.  Instead, we will consider the effect of a heavy scalar $S$ or a heavy fermion $\psi$ to represent new physics.  We couple the heavy scalar $S$ to $\phi$ by the addition of the potential
\begin{equation}
V(\phi,S) =\frac{1}{2} m_S^2 S^2 + \frac{1}{8} \lambda_S S^4 + \frac{1}{2} \eta_P \phi S^2 + \frac{1}{2} \lambda_P \phi^2 S^2\,.
\end{equation}
Alternatively, we couple the heavy fermion $\psi$ of mass $m_\psi$ by adding a Yukawa coupling of the form.
\begin{equation}
\mathcal{L}(\phi,\psi) = -y \phi \bar{\psi}\psi\,.
\end{equation}

We now decide on how to compute the effects of new physics on tunneling amplitudes.  Our strategy follows that of Weinberg \cite{Weinberg:1992ds}, wherein the heavy degrees of freedom are integrated out in matter analogous to the Born-Oppenheimer approximation 
\begin{gather}
Z = \int \mathcal{D} \phi\, \mathcal{D} S\, e^{i S[\phi, S]} = \int \mathcal{D} \phi\, e^{i W[\phi]}\,, \text{ or}\\
Z = \int \mathcal{D} \phi\, \mathcal{D} \psi\mathcal{D} \bar\psi\, e^{i S[\phi, \bar\psi, \psi]} = \int \mathcal{D} \phi\, e^{i W[\phi]}
\end{gather}
yielding an action functional $W[\phi]$ which is equal to the sum of connected diagrams with external $\phi$ lines and internal $S$ or $\psi$ lines.   The partial tunneling amplitude will subsequently be evaluated based on $W[\phi]$ as outlined in the previous section.  However $W[\phi]$ is a complicated nonlocal functional of $\phi(x)$, and evaluating it for an arbitrary profile as in (\ref{eq:gaussianfamily}) is impossible.  However, as Weinberg argues, a tractable approximation can be made based on the coupling constant expansion if the quartic self coupling $\lambda_\phi$ and the coupling to new physics $\lambda_\text{portal} = \{\lambda_P \text{ or } y^2\}$ satisfy the relationship
\begin{equation}
\lambda_\phi^2 \sim \lambda_\text{portal}\,,
\end{equation}
similar to the one used to analyze the Coleman-Weinberg mechanism\cite{Coleman:1973jx}.  In that case, the leading contribution in the coupling constant expansion is the one loop effective potential (with only new physics integrated out) evaluated at the Gaussian bubble
\begin{equation}
W[\phi_G] = \int d^4 x \big[\frac{1}{2}\partial_\mu \phi \partial^\mu \phi - V^\text{1-loop}_\text{eff}(\phi_\text{G}(x)) + \ldots \big] \,
\end{equation}  
and will be the order to which all subsequent calculations are accurate.  We emphasize that apart from the coupling constant expansion, we do not make any further approximations.  Retaining just the first few terms in the inverse mass expansion is inconsistent since the field strengths in the bubbles may be large.

Finally, since the low energy constants determine the measured masses and couplings of scalar quanta in the metastable point, we will work in the effective potential scheme where the renormalized parameters satisfy
\begin{gather}\label{eq:EffPotScheme}
V''_\text{eff}(0) = m_\phi^2\,,\enspace V'''_\text{eff}(0) = -6\eta_\phi\,, \enspace V''''_\text{eff}(0) = 3\lambda_\phi
\end{gather}
to prevent them from being modified upon the addition of new physics.

\section{Asymmetric double well potential}
We begin our analysis for the asymmetric double well potential given in (\ref{eq:lowEnergyPot}), with positive low energy constants $m_\phi^2$, $\eta_\phi$ and $\lambda_\phi$.  Throughout this and the next section we work in units normalized by the $\phi$ mass, so that $m_\phi=1$ and all other dimensional parameters are quoted in units of $m_\phi$.  Furthermore in this section, we fix the model parameters to be $\eta_\phi=0.25$, $\lambda_\phi=0.01$, $m_S=15$, $\eta_P=0.25$, $\lambda_P=1$, $m_\psi=15$ and $y=0.8$.

\begin{figure}[t]
\includegraphics[width=1.\columnwidth]{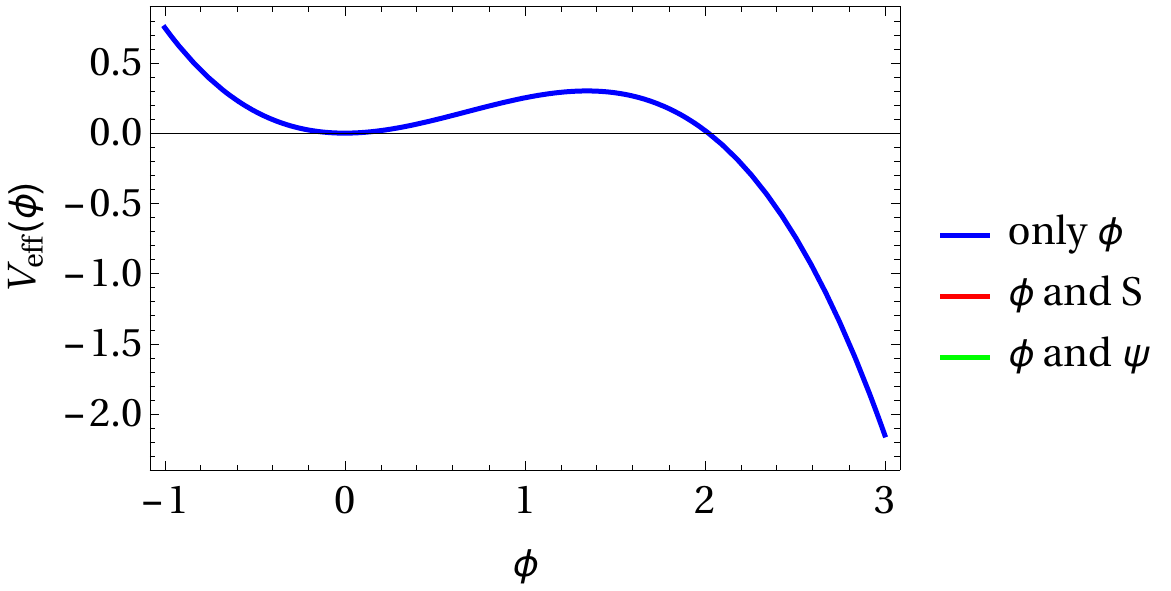}
\includegraphics[width=1.\columnwidth]{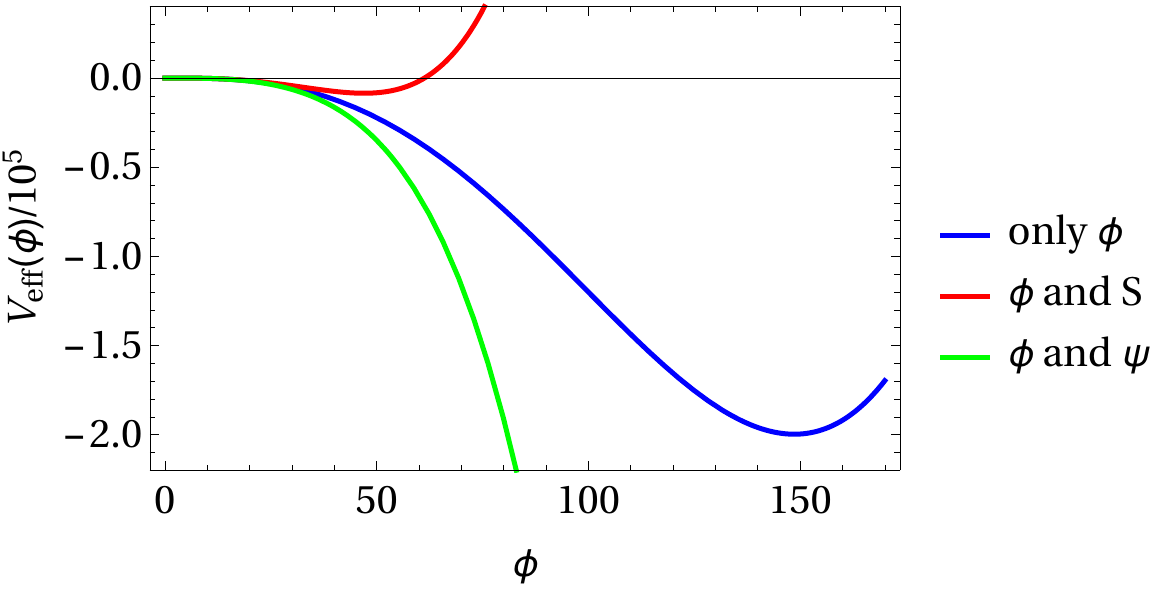}
\caption{The double well potential for $\phi$ alone in blue, with scalar $S$ in red, and fermion $\psi$ in green.  The upper panel shows the potentials for small field values, and are visually indistinguishable.}\label{fig:pot}
\end{figure}

The tree-level potential (corresponding to no new physics) and the one-loop effective potentials (with only $S$ or $\psi$ integrated out) are displayed in Fig. \ref{fig:pot}.  Since they are evaluated in the effective potential scheme (\ref{eq:EffPotScheme}), the potential near the metastable point (upper panel) remains unaffected by the addition of new physics.  However, at larger field values (lower panel) the effect of new physics is apparent.

Following the method in Sec. \ref{sec:PartialAmplitude}, we calculate the partial amplitude of the false vacuum decay at $\phi=0$ into a final state bubble of the form in (\ref{eq:finalbubble}).  The requirement of energy conservation fixes the relationship between the bubble size $R$ and amplitude $f_\text{f}$.  In the absence of new physics, this relationship is determined by the tree level action and is given by 
\begin{equation}
R^2(f_\text{f}) =\frac{216 \sqrt{2}}{-72\sqrt{2} m_\phi^2 + 64\sqrt{3} \eta_\phi f_\text{f} - 9 \lambda_\phi f_\text{f}^2}  \,.
\end{equation}
The relationship in the presence of new physics requires the effective potential and we determine it numerically.\

\begin{figure}[t]
\includegraphics[width=1.\columnwidth]{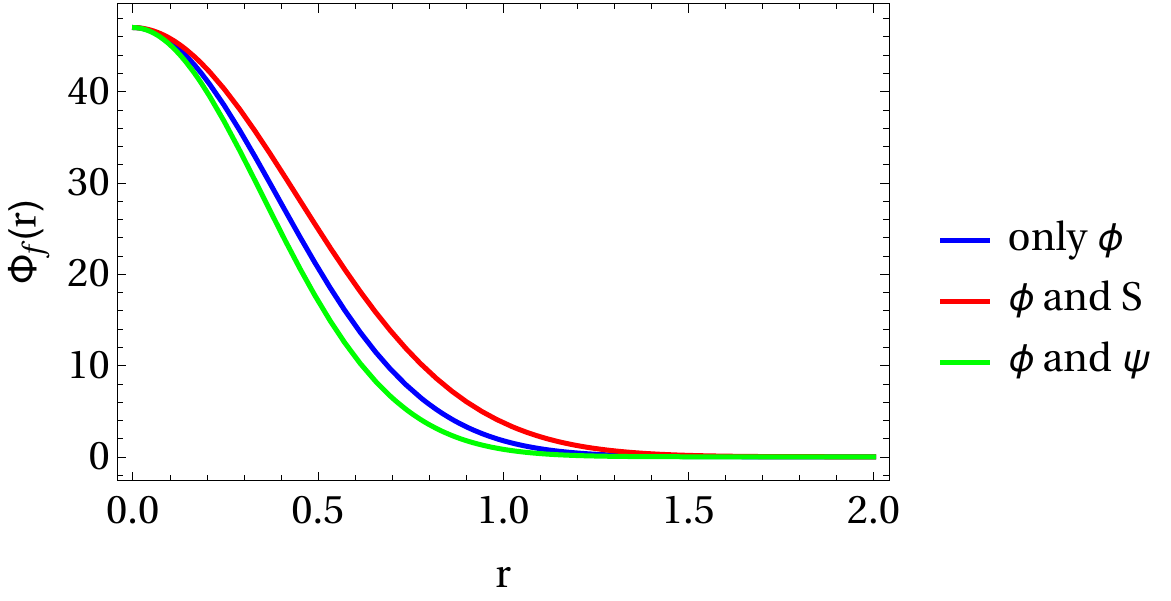}
\caption{The final state bubbles for $f_\text{f}=47$ without new physics in blue, and with new physics scalar $S$ in red, fermion $\psi$ in green.}\label{fig:bubble}
\end{figure}

In Fig. \ref{fig:bubble} we show the profile for a final state bubble with and without new physics for a fixed value of the field at the center of $f_\text{f} = 47$.  Although the final states are not exactly the same, we see that there is a characteristic scale $f_\text{f} \sim R^{-1}$ associated with the final state bubbles.  The addition of a boson $S$ stabilizes the effective potential.  Therefore to maintain energy conservation, the bubble must have a larger radius.  The addition of a fermion $\psi$ has the opposite effect, forcing a smaller bubble.  Note that bubbles with $f_\text{f}$ much larger than the true minimum are not possible.  Furthermore, since the scalar $S$ stabilizes the potential bringing the minimum to lower field values, some bubbles which were previously possible are no longer available as final states.

\begin{figure}[h]
\includegraphics[width=1.\columnwidth]{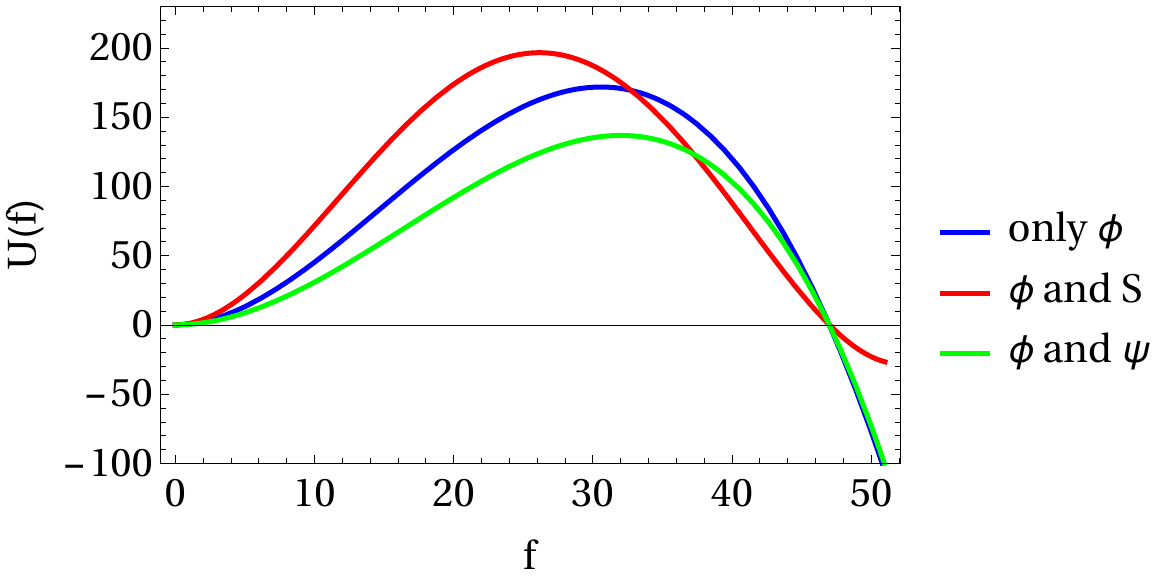}
\includegraphics[width=1.\columnwidth]{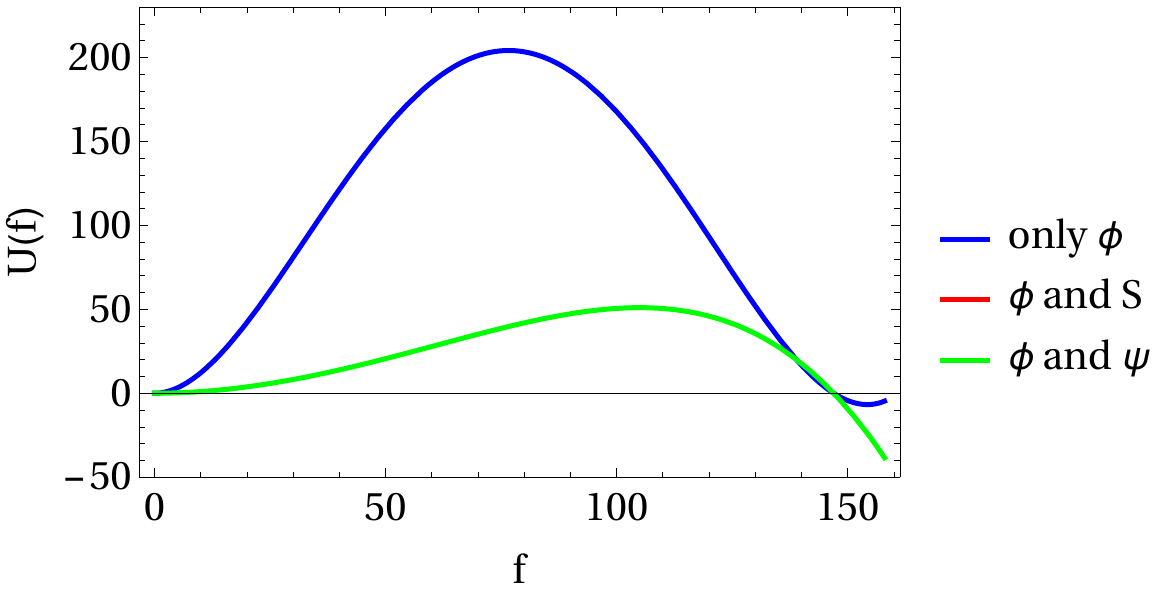}
\caption{Reduced potential $U(f)$ which a final state bubble has to tunnel through. Upper panel: $f_\text{f}=47$, lower panel: $f_\text{f}=147$.  Without new physics in blue, and with new physics scalar $S$ in red and with fermion $\psi$ in green.  Note that the stabilizing effect of the scalar $S$ has shut down the decay channel to an $f_\text{f}=147$ bubble.}\label{fig:U}
\end{figure}

We proceed to evaluate the partial amplitude by reducing the field theory problem to a quantum mechanical problem by restricting the path integral to the family of Gaussian bubbles given in (\ref{eq:gaussianfamily}).  Without new physics, reduced quantum mechanical potential $U(f)$ for the dynamical variable $f(t)$ is
\begin{equation}
U(f)=\frac{\pi^3 R^6}{16} \Big( \frac{3}{R^2} f^2 + m_\phi^2 f^2 - \frac{8\eta_\phi}{3\sqrt{6}} f^3  + \frac{\lambda_\phi}{8\sqrt{2}} f^4 \Big)\,.
\end{equation}
With new physics, there is an additional contribution from the effective potential, which we evaluate numerically.  We show the reduced potential $U$ the system must tunnel through in Fig. \ref{fig:U} for two representative final state bubbles, $f_\text{f}=47$ and $f_\text{f}=147$. Observe that new physics significantly changes this potential for final state bubbles which are smaller than the scale set by new physics at $m_S=m_\psi=15$. 

\begin{figure}[h]
\includegraphics[width=1.\columnwidth]{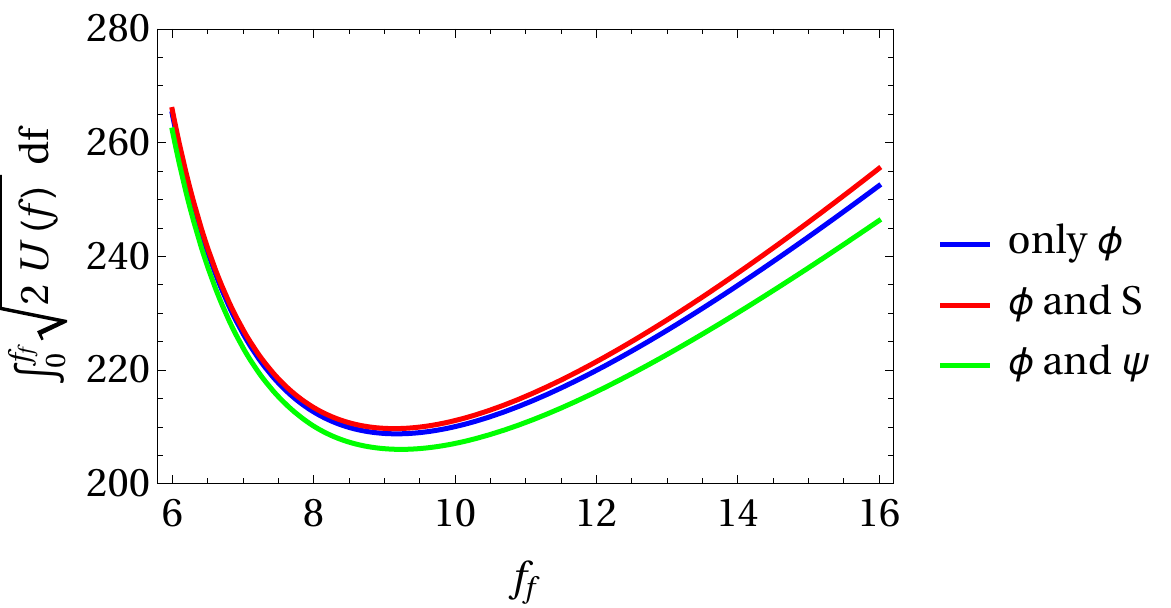}
\includegraphics[width=1.\columnwidth]{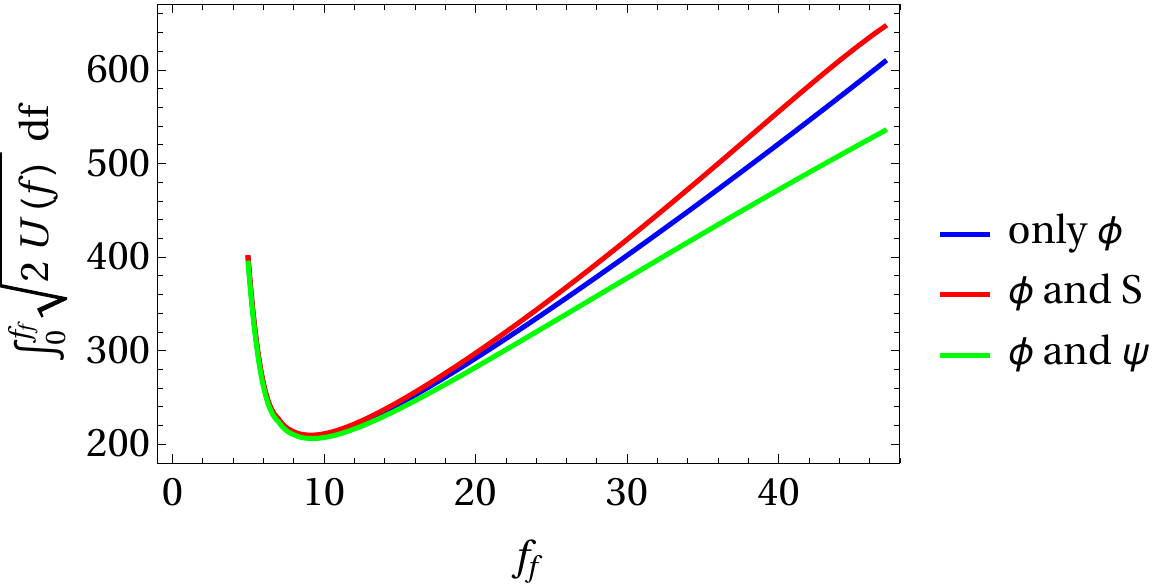}
\includegraphics[width=1.\columnwidth]{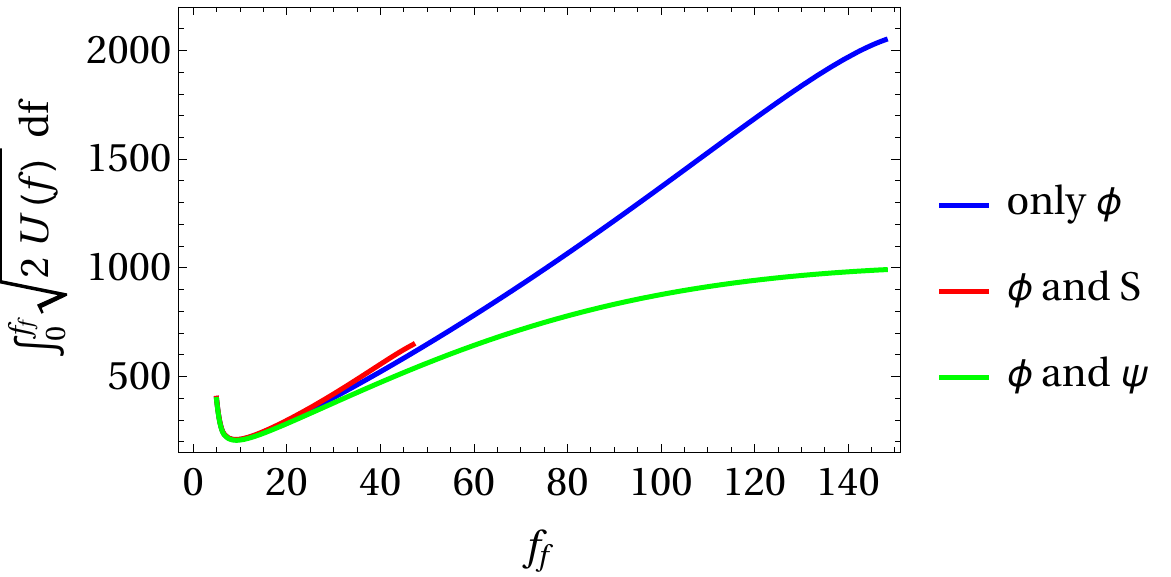}
\caption{WKB exponent for different final state bubbles, characterized by scale $f_\text{f}$ without new physics in blue, and with new physics scalar $S$ in red, and fermion $\psi$ in green.}\label{fig:SE}
\end{figure}

In Fig. \ref{fig:SE} we plot the WKB exponent $\int_{0}^{f_\text{f}} \sqrt{2 U(f)} df$ which controls the partial tunneling rate as a function of $f_\text{f}$ characterizing the scale of the final state bubble.  The minimum of WKB exponent at a low scale of $f_\text{f}^\text{crit} \approx 9$ corresponds to a close approximation of the Coleman-Callan bounce which dominates the total rate.  Notice that for bubbles of smaller radius (large $f_\text{f}$), the WKB exponent is greatly modified by the presence of new physics.  But, the amplitude for the decay into the dominant final state bubble $f_\text{f}^\text{crit}$ remains relatively unaffected.  Therefore, in this model, the total metastable decay rate (summed over all final states) will remain unaffected.  Although not displayed here, we have numerically confirmed that new physics decouples from the low energy WKB exponent like  $m_S^{-2}$ or $m_\psi^{-2}$. 

\section{Unbounded Potential}
\begin{figure}[t]
\includegraphics[width=1.\columnwidth]{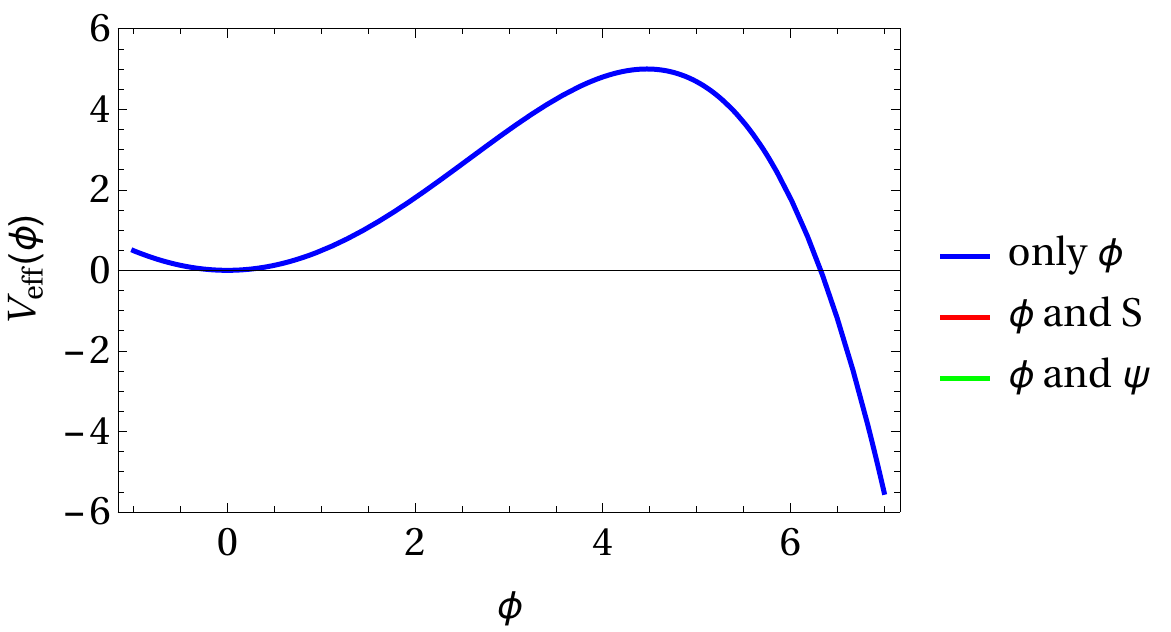}
\includegraphics[width=1.\columnwidth]{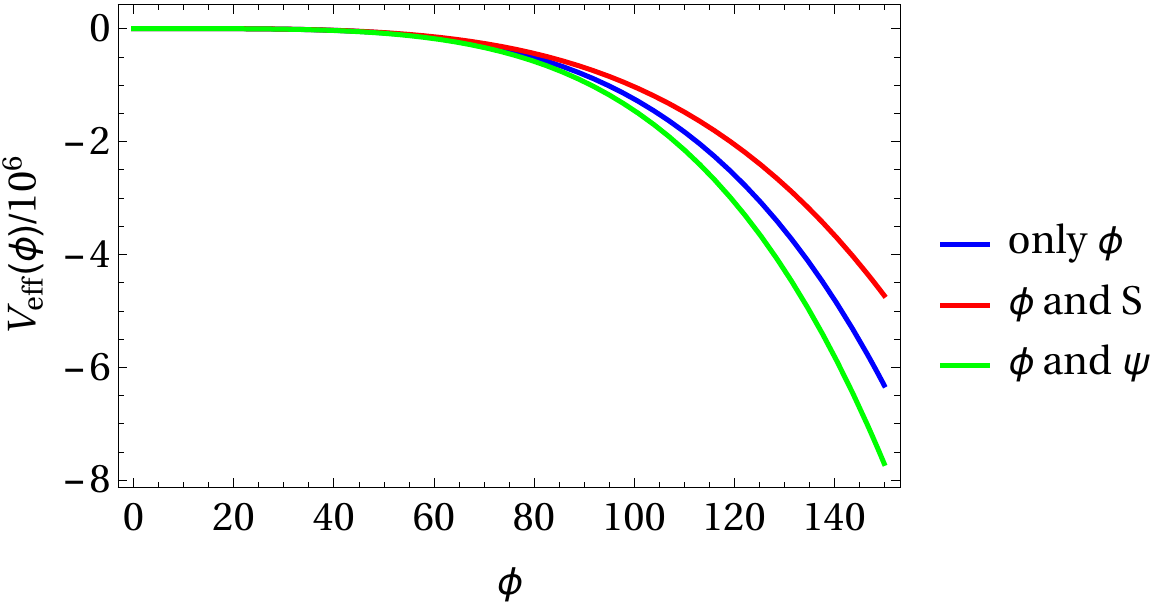}
\caption{The unbounded potential for $\phi$ alone in blue, with the scalar $S$ in red, and with the fermion $\psi$ in green.  The upper panel shows the potentials for small field values, and are visually indistinguishable.}\label{fig:pot2}
\end{figure}

We now turn to the case of the unbounded potential where we take (\ref{eq:lowEnergyPot}) with $\lambda_\phi$ negative and small, and for simplicity $\eta_\phi=0$.  This is similar to the Standard Model potential for high field values.  For our numerical study, we fix the model parameters to be $m_\phi=1$, $\eta_\phi=0$, $\lambda_\phi=-0.1$, $m_S=30$, $\eta_P=0$, $\lambda_P=1$, $m_\psi=30$ and $y=0.8$.  We display the form of the tree-level potential (no new physics) and one-loop effective potentials (with $S$ or $\psi$ integrated out) in Fig. \ref{fig:pot2}.  As before, since the renormalized parameters are defined in the effective potential scheme, the shape of the potential remains unchanged near the location of the metastable vacuum.

Before we continue, we remind the reader the situation for this theory in the absence of new physics \cite{Affleck:1980mp}.  The total rate is conventionally determined by solving for the field configuration (bounce) that minimizes the Euclidean action.  However, by simple scaling arguments, one can show no such solution is to be found since a lower action can be obtained for smaller bounces.  However, a limiting value of the action exists and can be extracted by temporarily imposing a constraint
\begin{equation} 
\int d^4 x \phi^n(x) = \rho^{4-n} \,,
\end{equation}
to the bounce.  This allows one to solve for the minimum for the action, corresponding to the constrained bounce.  The result can be inserted in the action, and upon taking the limit $\rho\rightarrow0$, the limiting value of 
\begin{equation}\label{eq:constrainedAction}
S = \frac{8 \pi^2}{3 \vert\lambda_\phi\vert} \,
\end{equation}
is obtained.  

What does this imply when new physics is added to the model?  Since the dominant contribution to the tunneling amplitude comes from a narrow configuration with an infinite field strength at the center, we expect amplitudes to small Gaussian bubbles (large $f_\text{f}$) to be significantly modified.  We confirm this expectation below by evaluating the partial amplitudes to Gaussian bubbles as outlined in Sec \ref{sec:PartialAmplitude}.

\begin{figure}[t]
\includegraphics[width=1.\columnwidth]{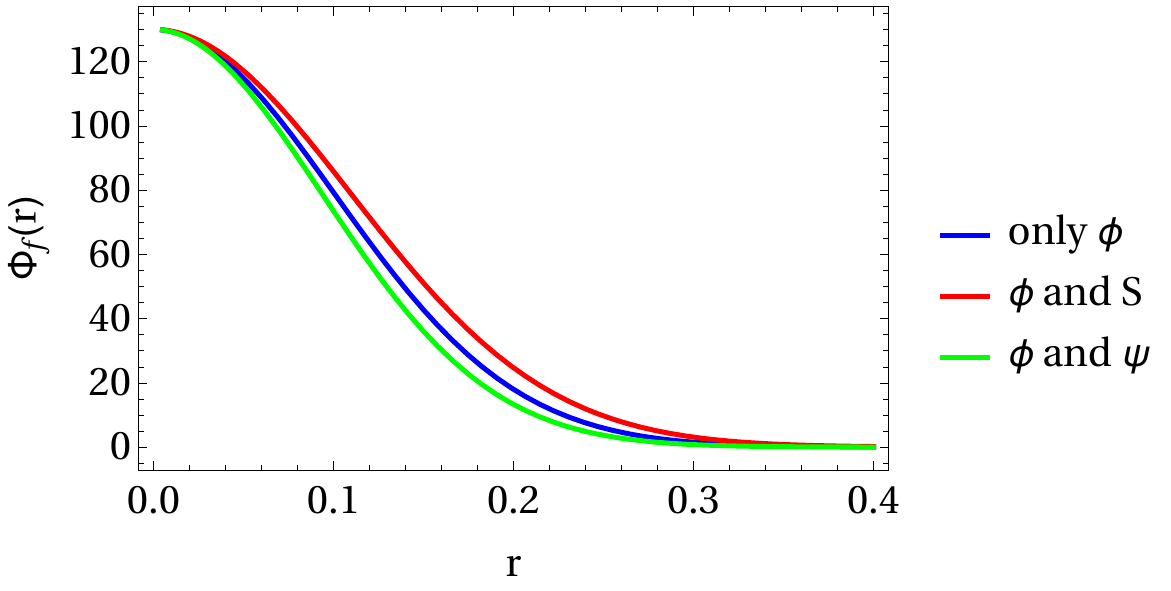}
\caption{The final state bubble with $f_\text{f}=100$ with only scalar $\phi$ in blue and with new physics scalar $S$ in red and fermion $\psi$ in green.}\label{fig:bubble2}
\end{figure}

\begin{figure}[t]
\includegraphics[width=1.\columnwidth]{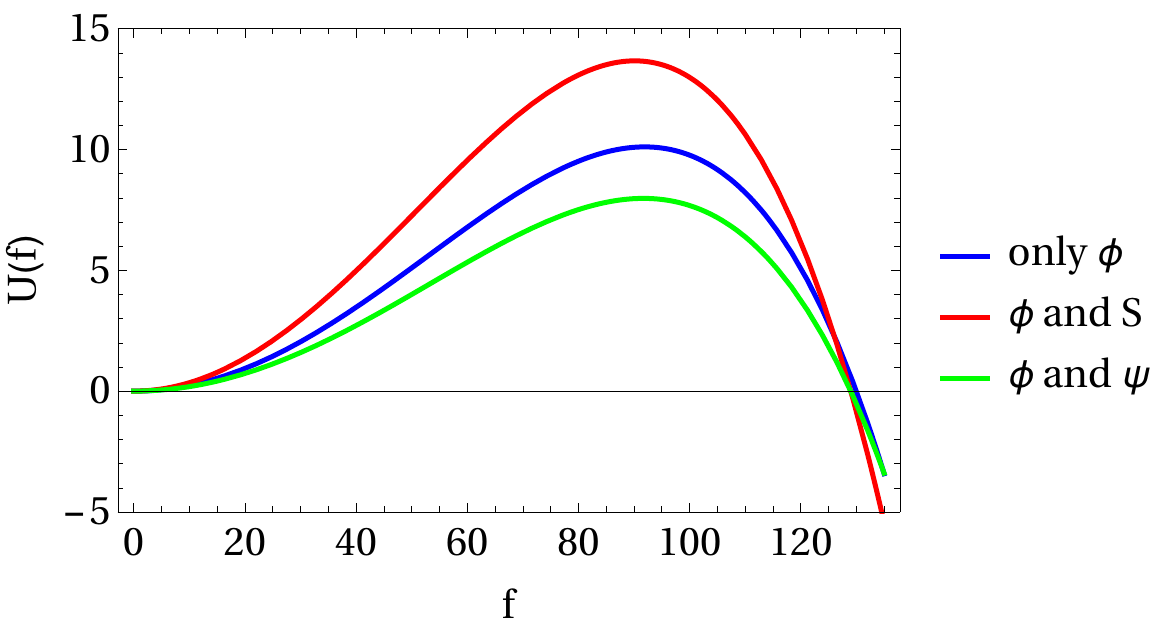}
\caption{Reduced potential $U(f)$ which a final state bubble with $f_\text{f}=100$ has to tunnel through in blue, with scalar $S$ in red and with fermion $\psi$ in green.}\label{fig:U2}
\end{figure}

In Fig. \ref{fig:bubble2} we show an example final state Gaussian bubble with and without new physics for fixed field value of $f_\text{f}=100$ inside the bubble.  In Fig. \ref{fig:U2} we display the reduced potential $U(f)$ which the system must tunnel through to reach the final state bubble.  As in the case of the asymmetric double well, new physics makes a substantial modification to the reduced potential since the size of the chosen final state bubble is much smaller than the scale set by new physics ($m_S=m_\psi=30$).
\begin{figure}[t]
\includegraphics[width=1.\columnwidth]{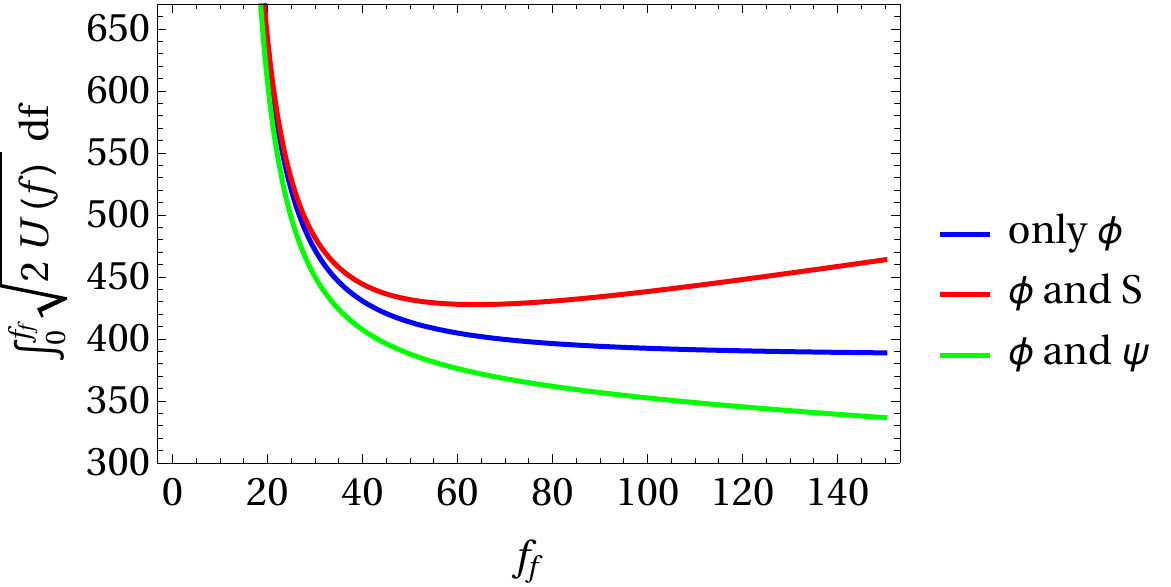}
\caption{WKB exponent for different final state bubbles for only $\phi$ in blue, with $S$ in red and with $\psi$ in green.}\label{fig:SE2}
\end{figure}

In Fig. \ref{fig:SE2} we plot the WKB exponent $\int_{0}^{f_\text{f}} \sqrt{2 U(f)} df$ which controls the partial tunneling rate as a function of $f_\text{f}$ characterizing the scale of the final state bubble.  The blue curve corresponds to the low energy theory, and is monotonically decreasing without exhibiting a local minimum.  This is a reflection of the absence of a stationary solution, and where the limiting value as $f_\text{f}\rightarrow\infty$ closely approximates the value given in (\ref{eq:constrainedAction}) as determined by constrained bounce.

The addition of new physics significantly modifies the amplitudes corresponding to final states of small bubbles which we now elaborate.  The red curve in Fig. \ref{fig:SE2} is the result of adding the heavy scalar $S$.  The WKB exponent has a local minimum corresponding to a critical bubble $f_\text{f}^\text{crit}$ due to the presence of a new stabilizing scale $m_S$, and gives the dominant contribution to the total width of the metastable vacuum.  Furthermore, the WKB exponent for bubbles whose inverse radius is larger than the scale of new physics have been significantly modified, exhibiting the expected sensitivity to new physics.

One might wonder how the critical bubble and the associated WKB exponent behaves as $m_S$ is increased.  We can find the behavior by first retaining the leading term of the large $m_S$ expansion of the one loop effective potential to construct the action functional
\begin{multline}\label{eq:asymW}
W[\phi] = \int d^4 x \big[\frac{1}{2}\partial_\mu \phi \partial^\mu \phi - V(\phi) \\
- \frac{1}{64\pi^2} \frac{\lambda_P^3}{3} \frac{\phi^6}{m_S^2} + \mathcal{O}\big(\frac{1}{m_S^4}\big)\big].
\end{multline}
Then the asymptotic behavior of the WKB exponent follows, which allows us to find the asymptotic behavior of the critical Gaussian bubble, its size, and the critical WKB exponents
\begin{equation}\label{eq:critBubb} 
f^\text{crit}_\text{f} \sim  a\frac{1}{\lambda_P^{3/4}} \sqrt{m_\phi m_S}\,,
\end{equation}
\begin{equation} 
R^\text{crit} \sim b \frac{1}{\sqrt{\lambda_\phi}} (f^\text{crit}_\text{f})^{-1}\,,
\end{equation}
\begin{equation} \label{eq:critWKB}
\int_{0}^{f_\text{f}} \sqrt{2 U(f^\text{crit}_\text{f})} df \sim \frac{1}{\lambda_\phi} + c \frac{\lambda_P^{3/2}}{\lambda_\phi^2} \frac{m_\phi}{m_S}\,,
\end{equation}
as $m_S\rightarrow\infty$, with $a$, $b$, and $c$ positive.
Since $f_\text{f}^\text{crit}$ in (\ref{eq:critBubb}) only grows like $\sqrt{m_S}$, it is never able to reach the scale of new physics $m_S$, a posteriori justifying the approximation in (\ref{eq:asymW}).  Furthermore, as $m_S$ is raised, the scalar $S$ representing new physics decouples from the WKB exponent in (\ref{eq:critWKB}), and the critical bubble goes over to the massless case that is obtained by the method of constrained bounce.  We note that the scaling derived above may be different for other theories, such as if $\eta_\phi \neq 0$.  But because the scalar $S$ effectively stabilizes the potential, it must decouple from the full width as $m_S\rightarrow\infty$.

For fermionic new physics, the effect is the opposite.  For a heavy fermion $\psi$, the WKB exponent is given by the green curve in Fig. \ref{fig:SE2}.  The destabilizing effect of adding a fermion prevents a local minimum from developing.  As a result, the total width continues to be dominated by infinitesimally small bubbles with field strengths that lie far beyond the scale of new physics, but without a limiting value.   While the heavy fermion decouples from the partial amplitude as $m_\psi\rightarrow\infty$ for any given final state Gaussian bubble of fixed $f_\text{f}$, it does not decouple from the \emph{total} decay width.  We point out that the reason for the drastic change in the total width is due to the unboundedness of the low energy potential $V(\phi)$ with $\lambda < 0$.

We have not resummed large logarithms through the renormalization group equations (RGEs).  While its inclusion can quantitatively change the impact of new physics on the total width, our point concerning the decoupling of new physics from partial amplitudes is unchanged.  This is because for final state bubbles with an inverse radius smaller than the scale of new physics, the running of coupling constants are induced by RGEs with beta functions appropriate only to low energy physics.  The effect of new physics on the WKB exponent will continue to be non-logarithmic as in (\ref{eq:critWKB}), and will decouple from the amplitude.  However, for inverse radii larger than the scale of new physics, the beta function is altered, causing a sizeable change in the WKB exponent.

We close this section with a few remarks concerning the implications of our findings on the vacuum instability in the standard model.  In the standard model the inverse radius of the dominant bounce is $10^{17}$ GeV.   In this regime, the scalar potential is well approximated by the quartic term, making it similar the case of the ``unbounded potential'' studied above.  Our findings suggest that adding a fermionic degree of freedom above that scale with a sufficiently large coupling would lead to a large change in the total width of the vacuum, confirming the original observation by Branchina.  However, our result does not suffer from the breakdown of the effective theory.

\section{Discussion and Summary}
In this paper, we numerically studied the impact new physics at the high scale may have on vacuum metastability in the $\phi^4$ theory without resorting to an effective theory description which is liable to break down.   We showed that some form of the decoupling theorem applies to partial amplitudes for decay processes to specific final state bubbles of a characteristic size.  Amplitudes for decay to final state bubbles of inverse radii larger than the scale of new physics can be significantly modified by the addition of new physics, while those for bubbles of smaller inverse radii are insensitive to new physics.  Because the total lifetime is given by the sum over partial rates for all possible final states, the inclusion of new physics may have the paradoxic effect of significantly altering the lifetime due to its effect on bubbles of large inverse radii.  Our findings suggest that the addition of scalar degrees of freedom has a stabilizing effect, and therefore decouples from the total lifetime.  But the addition of fermionic degrees of freedom with large Yukawa couplings can destabilize the system to the extent that its effect does not decouple.

\acknowledgements
BR is supported by the Alexander von Humboldt Foundation. BR acknowledges project 8799 by the Croatian Science Foundation.

\bibliography{tunneldec}

\end{document}